\documentstyle[prl,aps]{revtex}
\begin{document}

\twocolumn[\hsize\textwidth\columnwidth\hsize\csname
@twocolumnfalse\endcsname

\title{Plasmon excitation by charged particles interacting with metal
surfaces}
\author{A. Bergara$^1$, J. M. Pitarke$^{1,2}$, and R. H. Ritchie$^3$}
\address{$^1$Materia Kondentsatuaren Fisika Saila, Zientzi Fakultatea,
Euskal Herriko Unibertsitatea,\\ 
644 Posta Kutxatila, 48080 Bilbo, Basque Country, Spain\\
$^2$Donostia International Physics Center (DIPC) and Centro Mixto
CSIC-UPV/EHU\\ 
$^3$Oak Ridge National Laboratory, P.O. Box 2008, Oak Ridge, Tennessee
37831-6123}
\date{\today}
\maketitle

\begin{abstract}
Recent experiments (R. A. Baragiola and C. A. Dukes, Phys. Rev. Lett. {\bf
76},
2547 (1996)) with slow ions incident at grazing angle on metal surfaces
have shown that bulk plasmons are excited under conditions where the
ions do not penetrate the surface, contrary to the usual statement that probes
exterior to an electron gas do not couple to the bulk plasmon. We here use the
quantized hydrodynamic model of the bounded electron gas to derive an explicit
expression for the probability of bulk plasmon excitation by external
charged particles moving parallel to the surface. Our results indicate that
for
each ${\bf q}$ (the surface plasmon wave vector) there exists a
continuum of bulk plasmon excitations, which we also observe within the
semi-classical infinite-barrier (SCIB) model of the surface.
\end{abstract}

]

It is well known that charged particles interacting with solid surfaces can
create electronic collective excitations in the solid. These are
bulk\cite{Pines} and
surface\cite{Ritchie57} plasmons. In the absence of electron-gas
dispersion, the
scalar electric potential due to bulk plasmons vanishes outside the
surface\cite{Mahan};
hence, in this case probes exterior to the solid can only generate surface
excitations. That electron-gas dispersion allows external probes to interact
with bulk plasmons was discussed by Barton\cite{Barton} and
Eguiluz\cite{Eguiluz2}, and more recently by Nazarov {\it et
al}\,\cite{Nazarov}. Nevertheless, the fact that within a non-local
description
of screening bulk plasmons do give rise to a potential outside the solid
has been
ignored over the years\cite{Gersten,Deabajo}, even when electron-gas
dispersion has been included; also, current non-local theories of plasmon
excitation by external charges have not shown evidence for bulk plasmon
excitation outside the solid\cite{Deabajo,Gaspar,Monreal}. Recently, Baragiola
and Dukes\cite{Baragiola} have studied the emission spectra produced by
slow ions
that were incident at grazing angle; their data indicate that the bulk
plasmon is
importantly involved in the emission process, though the projectiles are not
expected to have penetrated into the solid. Bulk plasmon excitation in
electron
emission spectra produced by slow multiply charged ions has also been
investigated\cite{Stolterfoht}, with projectiles that may enter       the
solid.

In this letter we derive, within the quantized hydrodynamic model of the
bounded electron gas\cite{Wilems}, an explicit expression for the
probability of
bulk plasmon excitation by external charged particles moving parallel to a
jellium surface. Our model, which assumes a sharp electron density profile
at the
surface, neatly displays the role of bulk plasmon excitations in
the interaction of charged particles moving near a metal surface. We also
demonstrate that our results for the total energy-loss probability agree
with standard calculations\cite{Nunez} derived either by solving the
linearized
Bloch hydrodynamic equations\cite{Ritchiet} or within the semi-classical
infinite-barrier (SCIB) model of the surface\cite{Ritchiem,Griffin} with the
hydrodynamic approximation for the bulk dielectric function. Though it has
been
generally believed that when charged particles move outside the solid the
energy
loss predicted in these models is fully described by the excitation of surface
plasmons\cite{Deabajo}, we demonstrate that for each ${\bf q}$ (the surface
plasmon wave vector) excitation of both a discrete surface plasmon and a
continuum of bulk plasmons contribute to the total energy-loss
probability, in agreement
with the prediction of our quantized hydrodynamic scheme.

Take an inhomogeneous electron system capable of self-oscillations about
a ground state described by density-functional theory (DFT)\cite{Kohn}. In the
hydrodynamic limit\cite{Theory} the system is characterized by the electron
density
$n({\bf r},t)$ and a velocity field ${\bf u}({\bf r},t)$. The total energy of
the system can then be expressed as\cite{Ying} (we use atomic units throughout,
i.e., $e^2=\hbar=m_e=1$)
\begin{eqnarray}
H=&&G\left[n({\bf r},t)\right]+\int d{\bf r}n({\bf
r},t)\left[{1\over2}|{\bf\nabla}\psi|^2-V_0-V_1\right]\cr\cr
&&+{1\over 2}\int
d{\bf r}d{\bf r}'{n({\bf r},t)n({\bf r}',t)\over|{\bf r}-{\bf r}'|},
\end{eqnarray}
where irrotational flow has been assumed, i.e.,
${\bf u}({\bf r},t)=-{\bf\nabla}\psi$,
and retardation effects have been neglected. $G\left[n({\bf r},t)\right]$
represents the exchange, correlation and internal kinetic energies of the
electron system. $V_0$ is the electrostatic potential due to the
neutralizing ionic background of density $n_0$, and $V_1$ represents the
external
perturbation. From Eq. (1) the basic hydrodynamic equations can be derived,
following Bloch's approach\cite{Theory}, and they can be linearized in the
deviation
$n-n_0$ from the equilibrium value to find the existence of self-sustaining
normal modes of oscillation. 

We consider a charged particle moving with velocity ${\bf v}$ outside of a
metallic surface, along a trajectory that is parallel to the surface, with a
classical charge distribution given at ${\bf r}=({\bf r}_\parallel,z)$ by
$\rho_{\rm ext}({\bf r},t)=Z_1\delta({\bf r}_\parallel-{\bf
v}t)\delta(z-z_0)$.
We represent the ionic background by a jellium model (the jellium occupying
the
half-space
$z<0$), and assume a sharp electron density profile at the surface. We also
neglect exchange-correlation contributions to
$G[n]$, and approximate it by the Thomas-Fermi functional\cite{Theory}. In
this
approximation, for each value of ${\bf q}$ (the wave vector parallel to the
surface) there exist both bulk and surface normal modes of oscillation with
frequencies given by the following dispersion relations\cite{Wilems}:
\begin{equation}
(\omega^B_{q,p})^2=\omega_{\rm p}^2+\beta^2(q^2+p^2)
\end{equation}
and
\begin{equation}
(\omega^S_{q})^2={1\over 2}\left[\omega_{\rm p}^2+\beta^2q^2+\beta
q(2\omega_{\rm p}^2 +\beta^2 q^2)^{1/2}\right],\label{B.3}
\end{equation}
respectively, where $\omega_{\rm p}=({4\pi n_0})^{1/2}$ is the so-called
plasma
frequency and $\beta$ represents the speed of propagation of hydrodynamic
disturbances in the electron system. We choose\cite{note1}
$\beta=\sqrt{3/5}q_F$, $q_F$ being the Fermi momentum. 

Now we follow Ref.\onlinecite{Wilems} to quantize, after linearization, the
hamiltonian of Eq. (1) on the basis of the normal modes corresponding to
Eqs. (2) and
(3), which we shall refer after quantization as bulk and surface plasmons,
respectively. We find
\begin{equation}
H=H_G+H_0^B+H_0^S+H_1^B+H_1^S,
\end{equation}
where $H_G$ represents the Thomas-Fermi ground state of the static unperturbed
electron system\cite{Theory}. $H_0^B$ and $H_0^S$ are free bulk and surface
plasmon hamiltonians:
\begin{equation}
H^B_0={1\over\Omega}\sum_{{\bf q},p>0}[1/2+\omega^B_{q,p}]\
a_{{\bf q},p}^{\dag}(t)\ a_{{\bf q},p}(t)\label{B.4}
\end{equation}
and
\begin{equation}
H^S_0={1\over A}\sum_{\bf q}\left[1/2+\omega^S_{q}\right]\
b_{{\bf q}}^{\dag}(t)\  
b_{{\bf q}}(t),\label{B.2}
\end{equation}
where $\Omega$ and $A$ represent the normalization volume and the
normalization
area of the surface, respectively, and where $a_{{\bf q},p}(t)$ and
$b_{{\bf q}}(t)$ are Bose-Einstein operators that annihilate bulk and
surface plasmons with wave vectors $({\bf q},p)$ and ${\bf q}$, respectively.
$H_1^{B/S}$ are contributions to the hamiltonian coming from the coupling
between the external particle and bulk/surface plasmon fields:
\begin{equation} 
H_1^{B/S}=\int{\rm d}{\bf r}\rho_{\rm ext}({\bf r},t)\phi^{B/S}({\bf
r},t),\label{59}  
\end{equation}
$\phi^{B/S}({\bf r},t)$ representing operators corresponding to the scalar
electric potential due to bulk/surface plasmons. Outside the metal
($z>0$),  
\begin{equation}
\phi^B({\bf r},t)=-{1\over\Omega}\sum_{{\bf q},p>0}f_{q,p}^B(z)
{\rm e}^{{\rm i}{\bf q}\cdot{\bf r}_\parallel}
\left[a_{{\bf q},p}^{\dag}(t)+a_{-{\bf q},p}(t)\right]\label{B.10}
\end{equation}
and
\begin{equation}
\phi^S({\bf r},t)=-{1\over A}\sum_{\bf q}\ f_{q}^S(z)
\ {\rm e}^{{\rm i}{\bf q}\cdot{\bf r}_\parallel}\
\left[b_{{\bf q}}^{\dag}(t)+b_{-{\bf q}}(t)\right],\label{B.7}
\end{equation}
where $f_{q,p}^B(z)$ and $f_q^S(z)$ are bulk
and surface coupling functions, respectively:
\begin{equation}
f_{q,p}^B(z)={\sqrt{2\pi/\omega^B_{q,p}}\,\omega_{\rm p}\,p\,{\rm
e}^{-qz}\over
\left[p^4+p^2(q^2+\omega_{\rm p}^2/\beta^2)+\omega_{\rm
p}^4/(4\beta^4)\right]^{1/2}}
\end{equation}
and 
\begin{equation}
f_{{\bf q}}^S(z)={\sqrt{\pi\gamma_q/\omega^S_{q}}\,\omega_{\rm p}\over
\left[q(q+2\gamma_q)\right]^{1/2}}\,{\rm e}^{-qz},
\end{equation}
and where $\gamma_q$ represents the so-called inverse decay length of  
surface plasmon charge fluctuations\cite{Eguiluz2}:
\begin{equation}
\gamma_q={1\over 2\beta}\left[-\beta
q+\sqrt{2\omega_{\rm p}^2+\beta^2q^2}\right].
\end{equation}
 
We derive the potential induced by the presence of the external perturbing
charge as the expectation value of the total scalar potential
operator\cite{Fetter}:
\begin{equation}
V^{ind}({\bf r},t)={<\Psi_0|\phi_H^B({\bf r},t)+\phi_H^S({\bf
r},t)|\Psi_0>\over<\Psi_0|\Psi_0>},
\end{equation}
where $|\Psi_0>$ is the Heisenberg ground state of the interacting system and
where $\phi_H^B({\bf r},t)$ and $\phi_H^S({\bf r},t)$ are the operators of
Eqs. (8) and (9) in the Heisenberg picture. Our results\cite{Bergara}
reproduce
previous calculations for the image potential [defined as half the induced
potential at the position of the charged particle that creates it] of a
static ($v=0$) external charged particle\cite{Eguiluz2}, which in the
case of
a non-dispersive electron gas ($\beta=0$) coincides with the classical image
potential\cite{Ritchiepl} $V_{\rm im}(z)=-(4z)^{-1}$.

The energy loss per unit path length of a moving charged particle can be
obtained
as the retarding force that the polarization charge distribution in the
electron
gas exerts on the projectile itself\cite{Echenique}:
\begin{equation} 
-{dE\over dx}={1\over v}\int{\rm d}{\bf r}\rho_{\rm ext}({\bf r},t){\bf
\nabla}V^{ind}({\bf r},t)\cdot{\bf v}.\label{59}  
\end{equation}
By introducing here the induced potential, which we evaluate
from Eq. (13) up to first order in $Z_1$, we find\cite{note2}
\begin{equation}
-{dE\over dx}={1\over
v}\int_0^{q_c}dq\int_0^{qv}
d\omega\,\omega\left[P^B_{q,\omega}+P^S_{q,\omega}\right],
\end{equation}
where $P^{B/S}_{q,\omega}$ represent probabilities per unit time, unit
wave number and unit frequency for the excitation of bulk/surface plasmons
with wave number $q$ and frequency $\omega$:
\begin{equation}
P^B_{q,\omega}=Z_1^2\,{q\over 2\pi^2}\int_0^\infty
dp{\omega_{q,p}^B\left[f_{q,p}^B(z_0)\right]^2\over\sqrt{q^2v^2-
\omega^2}}\delta(\omega-\omega^B_{q,p})
\end{equation}
and
\begin{equation}
P^S_{q,\omega}=Z_1^2\,{q\,\omega_q^S\left[f_q^S(z_0)\right]^2\over\pi\sqrt{q
^2v^2-
\omega^2}}\delta(\omega-\omega^S_{q}).
\end{equation}

For comparison, we note that by solving the linearized Bloch hydrodynamic
equations the total probability per unit time, unit wave number and unit
frequency for the external particle to transfer momentum $q$ and energy
$\omega$ to the electron gas is found as:
\begin{equation}
P_{q,\omega}=2Z_1^2\omega_{\rm p}^2\,{{\rm
Im}\left\{-1/\left[\omega_{\rm
p}^2-2\beta^2\Lambda_q(\Lambda_q+q)\right]\right\}\over\pi\sqrt{q^2v^2-
\omega^2}}{\rm e}^{-2qz_0},
\end{equation}
where
\begin{equation}
\Lambda_q={1\over\beta}\sqrt{\omega_{\rm p}^2+\beta^2 q^2-\omega(\omega+{\rm
i}\delta)},
\end{equation}
and where $\delta$ is a positive infinitesimal. The same result is
obtained\cite{Nunez} within either the specular reflexion
(SR)\cite{Ritchiem} or
the SCIB\cite{Griffin} model of the surface, as long as the hydrodynamic
approximation for the bulk dielectric function is considered.

Within the various semiclassical approaches leading to Eq. (18) the
role played by bulk and surface plasmons goes unnoticed. Furthermore, Eq. (18)
shows no evidence of losses at the bulk plasmon frequency, and it has been
generally believed that the energy loss predicted by this equation 
originates entirely in the excitation of surface plasmons. That this is not
the case is
clearly shown in Fig. 1. This figure shows $P_{q,\omega}$, as computed from
Eq.
(18) with $\omega_{\rm p}=15.8{\rm eV}$ corresponding to the bulk plasma
frequency of aluminum metal and with a finite damping parameter $\delta$
accounting for the finite lifetime of plasmon fields. The parallel momentum
transfer, the velocity and the distance of the particle trajectory above the
surface have been taken to be $q=0.4\,{\rm a.u.}$, $v=2\,{\rm a.u.}$ and
$z_0=1.0\,{\rm a.u.}$, respectively, and different values of
$\delta$ have been considered. One sees that loss occurs at the surface
plasmon energy $\omega_q^S$ given by Eq. (3), while a continuum of bulk
plasmon
excitations occurs at energies $\omega_{q,p}^B$ (see Eq. (2)), which all
are over $\omega^B_{q,0}=(\omega_{\rm p}^2+\beta^2q^2)^{1/2}$, as predicted by
Eqs. (16) and (17). In the limit as $\delta\to 0^+$, both bulk and surface
contributions to the total energy-loss probability of Eq. (18) exactly
coincide
with the predictions of Eqs. (16) and (17), thus demonstrating the full
equivalence between our quantized hydrodynamic scheme and the more standard
semiclassical approaches.  Also, either introduction of both Eqs. (16) and
(17)
into Eq. (15) or replacement of the integrand in Eq. (15) by the energy-loss
probability of Eq. (18) with $\delta\to 0$ result in exactly the same total
energy loss. As the damping parameter increases, surface plasmon excitation
broadens to energies over
$\omega_{q,0}^B$, and for $\delta\approx\omega_p/10$ bulk plasmon
contributions
to the energy loss go unnoticed.

Figure 2 exhibits, by a solid line, the energy loss per unit path length
versus the distance $z_0$ from the surface, as obtained from Eq. (15).  In this
case a proton ($Z_1=1$) moves with velocity $v=2\,{\rm a.u.}$ parallel to the
surface of a bounded electron gas of density equivalent to that of aluminum
($r_s=2.07$)\cite{note3}. Separate contributions from the excitation of
bulk and surface plasmons are represented by dashed and dotted lines,
as obtained with the use of Eqs. (16) and (17), respectively. One sees that
for $z_0<2\,{\rm a.u.}$ the contribution to the energy loss from the bulk channel
is important, while for larger values of $z_0$ the surface channel alone
gives a sufficiently accurate description of the total energy loss.

Figure 3 shows, by a solid line, results for the energy loss of Eq. (15), as a
function of the velocity of the projectile, with $Z_1=1$, $r_s=2.07$ and
$z_0=1\,{\rm a.u.}$. Dashed and dotted lines
represent separate contributions to the total energy loss from the
excitation of
bulk and surface plasmons, respectively. We note that for a projectile moving
at $z_0=1\,{\rm a.u.}$ the contribution to the energy loss from the bulk channel
is observable for all velocities, the relative importance of bulk plasmon
excitations becoming more important at velocities around the plasmon threshold
velocity when the projectile has enough energy to excite plasmons.
For comparison, the result one obtains either from Eq. (17) or Eq. (18) in the
case of a non-dispersive electron gas ($\beta=0$), which coincides with the
classical formula of Echenique and Pendry\cite{Pendry}, is represented by a
dashed-dotted line.

In conclusion, we have used the quantized hydrodynamic model of the
bounded electron gas to demonstrate that bulk plasmons undergo real
excitations, even in the case of charged particles that do not penetrate into
the solid. We have derived explicit expressions for the probability of
both bulk and surface plasmon excitation by external charged particles moving
parallel to the surface, which neatly display the role that bulk plasmon
excitation plays in the interaction of charged particles moving near a metal
surface. The full equivalence between our quantized hydrodynamic scheme and
the more standard semiclassical approaches has been demonstrated. It has
been generally believed that energy loss predicted by these approaches
originates entirely in the excitation of surface plasmons. However, we have
shown that for each value of the wave vector parallel to the surface both a
discrete surface plasmon excitation and a continuum of bulk plasmon excitations
contribute to the total energy-loss probability. We have also presented explicit
calculations of the energy loss per unit path length of protons moving outside
of a metal along a trajectory that is parallel to the surface, and our results
indicate that the contribution from the bulk channel is important for all
projectile velocities, as long as the distance from the surface is smaller
than a few atomic units.

We acknowledge partial support by the Basque Unibertsitate eta Ikerketa Saila
and the Spanish Ministerio de Educaci\'on y Cultura.

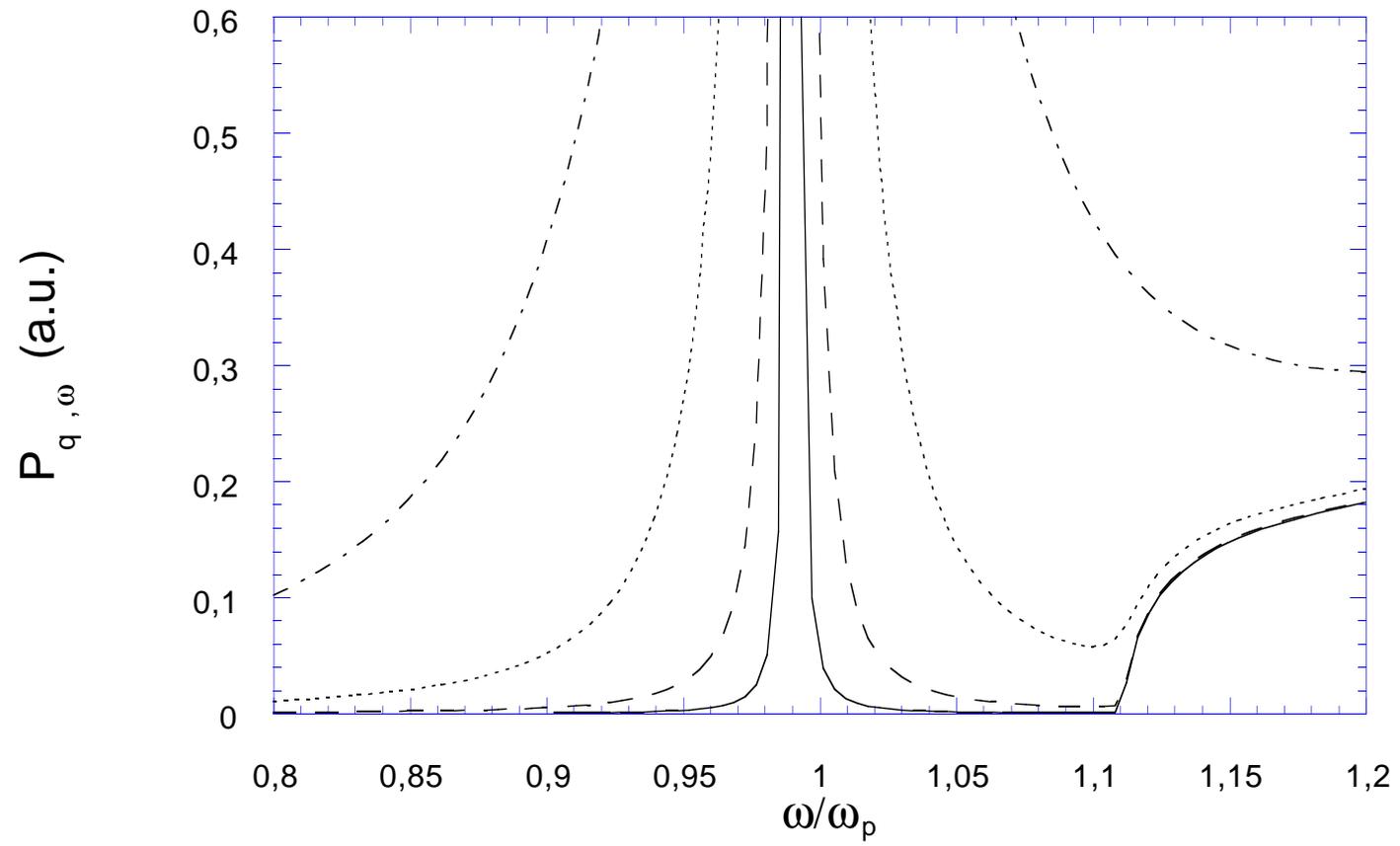
\begin{figure}
\caption{Energy-loss probability of Eq. (18), as a function of
$\omega/\omega_{\rm p}$, with $\omega_{\rm p}=15.8\,{\rm eV}$,
$q=0.4\,{\rm a.u.}$, $v=2\,{\rm a.u.}$,
$z_0=1\,{\rm a.u.}$, $Z_1=1$ and various values of the damping parameter
$\delta$: $10^{-4}\omega_{\rm p}$ (solid line), $10^{-3}\omega_{\rm p}$ (dashed
line), $10^{-2}\omega_{\rm p}$ (dotted line), and $10^{-1}\omega_{\rm p}$
(dashed-dotted line).} 
\end{figure}

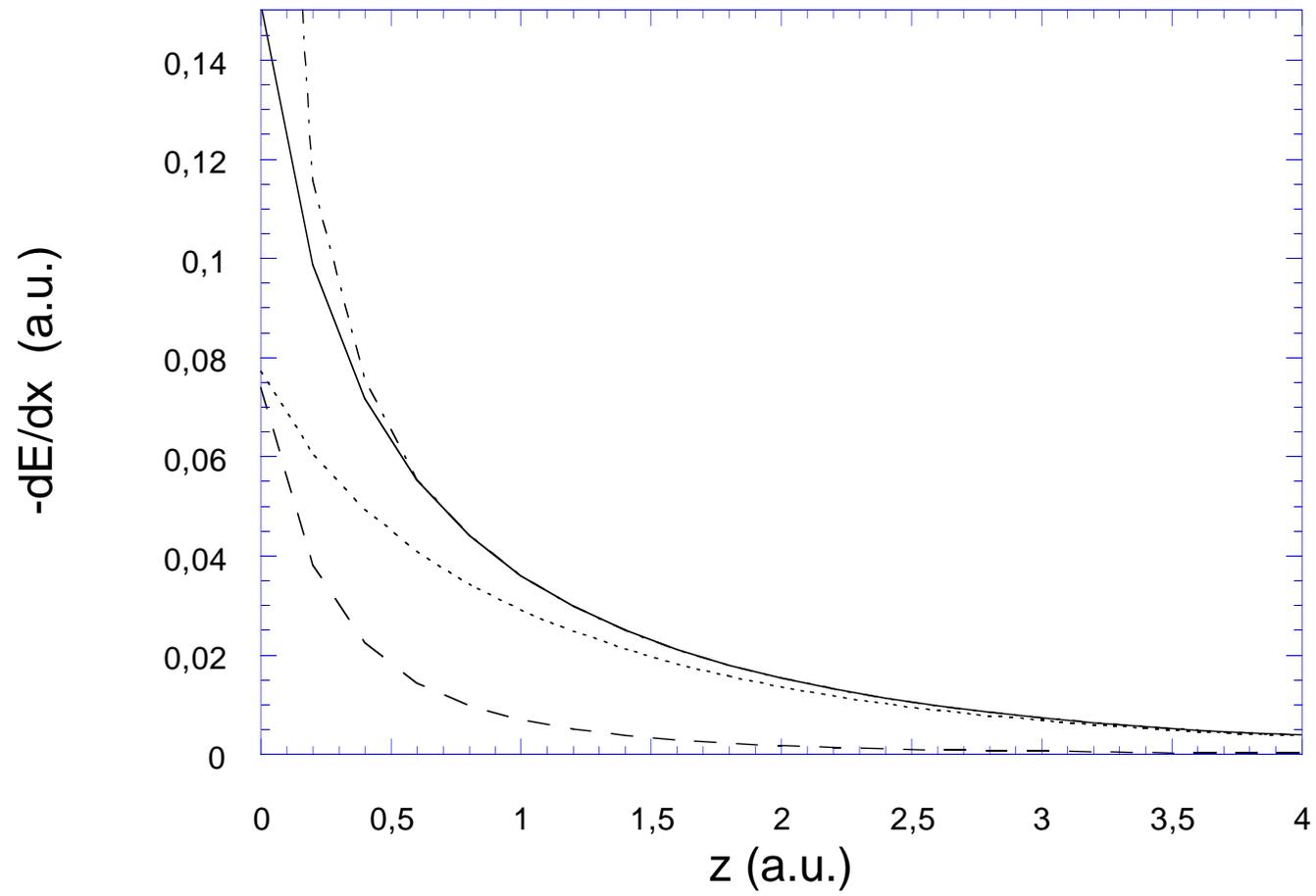
\begin{figure}
\caption{Energy loss per unit path length versus the distance from the surface
$z_0$, as obtained from Eq. (15) with either the use of Eqs. (16) and (17) or the
use of Eq. (18) and $\delta\to 0$ (solid line). Separate contributions from Eqs.
(16) and (17) are represented by dashed and dotted lines, respectively. A proton
($Z_1=1$) is assumed to move with velocity $v=2\,{\rm a.u.}$ parallel to the
surface of a bounded electron gas with an electron density equal to that of
aluminum ($r_s=2.07$). The dashed-dotted line represents the total
energy-loss obtained from Eq. (15) with no cut-off in the momentum integration
($q_c\to\infty$).}   
\end{figure}

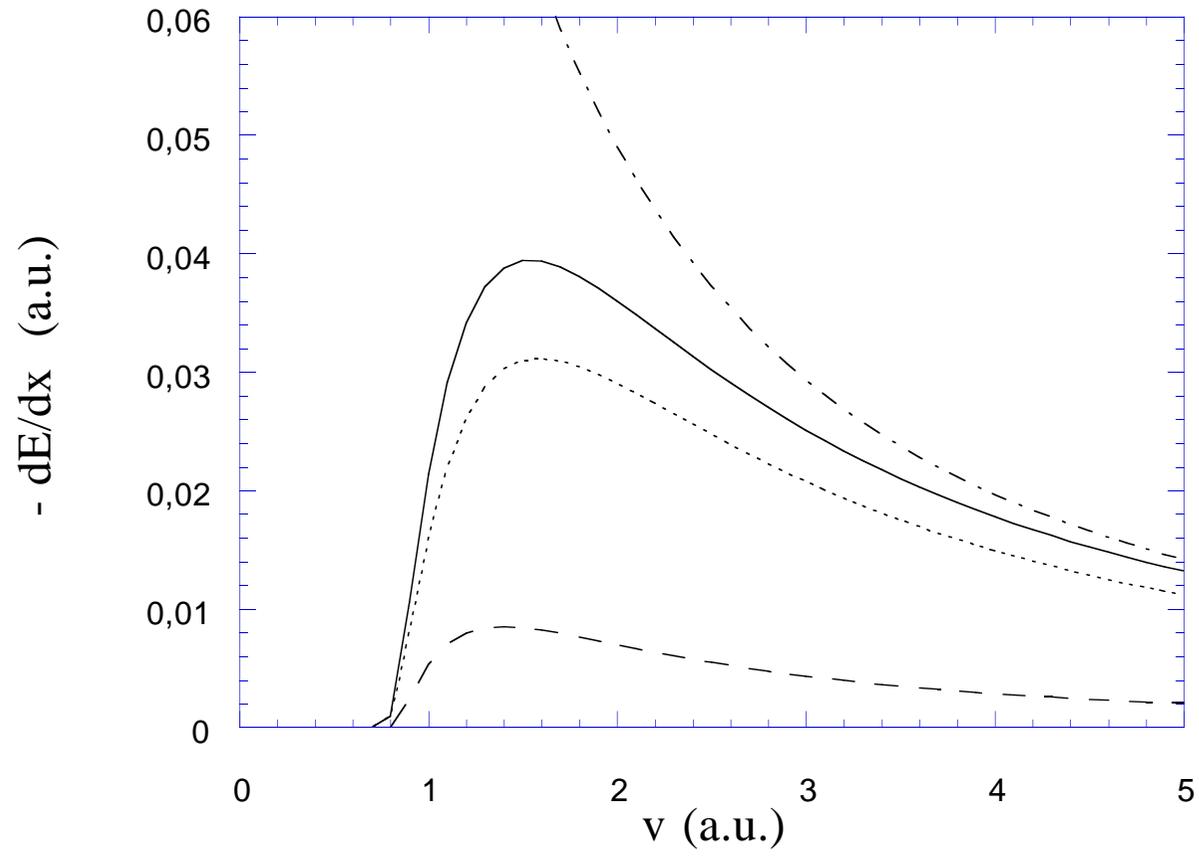
\begin{figure}
\caption{As in Fig. 2, but now the energy-loss per unit path length of Eq. (15)
is represented as a function of the velocity of the projectile for a given
value of the distance from the surface: $z_0=1\,{\rm a.u.}$. The dashed-dotted
line here represents the result obtained in the case of a
non-dispersive electron gas ($\beta=0$).}
\end{figure}


\begin{references}
\bibitem{note1} While use of the Thomas-Fermi $G[n]$ functional, which assumes
static screening, predicts $\beta=\sqrt{1/3}q_F$, the value
$\beta=\sqrt{3/5}q_F$ is expected to be appropriate when high frequencies of
the order of the plasma frequency are involved\cite{Pines}.
\bibitem{note2} The quantal hydrodynamic model of the electron gas assumes
that
infinitely long-lived plasmons are the only possible excitations for all wave
vectors. Since such an approximation is no longer valid for short wavelengths
where collective effects no longer occur, we use a cut-off in the momentum
integration, which we choose to be $q_c\approx 2v$.
\bibitem{note3} The so-called electron-density parameter $r_s$ is defined by
the relation $1/n_0=(4/3)\pi r_s^3$.
\end{references}
\end{document}